\documentclass[twocolumn,superscriptaddress]{revtex4}
 
\usepackage[german,english]{babel}
\usepackage{amssymb}
\usepackage{amsfonts}
\usepackage{amsmath}
 
\begin{document}
 
\title{Towards the molecular workshop: entropy-driven designer molecules,
entropy activation, and nanomechanical devices}

\date{\today}

\author{Andreas Hanke}
\affiliation{Department of Physics, Theoretical Physics, 1 Keble Road,
Oxford OX1 3NP, United Kingdom}
\affiliation{Department of Physics,
Massachusetts Institute of Technology, 77 Massachusetts Avenue, Cambridge,
Massachusetts 02139, USA}
\author{Ralf Metzler}
\affiliation{Department of Physics,
Massachusetts Institute of Technology, 77 Massachusetts Avenue, Cambridge,
Massachusetts 02139, USA}
\affiliation{NORDITA, Blegdamsvej 17, DK-2100 K{\o}benhavn {\O}, Denmark}

\begin{abstract}
We propose a new class of designer molecules with functional units
which are driven by {\em entropic} forces. 
This idea profits from the mechanically interlocked 
nature of topological molecules such as catenanes and rotaxanes, which 
allows for {\em mobile\/} elements whose accessible configuration space 
gives rise to entropic intramolecular forces. Such entropy-driven designer 
molecules open the possibility for externally controllable 
functional molecules and nanomechanical devices.
\end{abstract}

\maketitle

Complex molecules can be endowed with the distinct feature 
that they contain subunits which are linked to each other 
mechanically rather than chemically \cite{schill}. The investigation of 
the structure and properties of such interlocked {\em topological
molecules\/} is 
subject of the growing field of chemical topology \cite{frisch}; 
while speculations about the possibility of catenanes 
\footnote{{\em catena\/} (lat.), the chain.} (Olympic rings) date
back to the early 20th century lectures of Willst{\"a}tter, the 
actual synthesis of catenanes and rotaxanes 
\footnote{{\em rota\/} (lat.), the wheel; {\em axis\/} (lat.), 
the axle.} succeeded in 1958 \cite{schill}. Modern organic chemistry
has seen the development of refined synthesis methods to generate topological
molecules.

In parallel to the miniaturisation in electronics \cite{bishop} and the
possibility of manipulating single (bio)molecules \cite{strick}, 
supramolecular chemistry which makes use of chemical topology properties is
coming of age \cite{lehn,dietrich}. Thus, rotaxane-type molecules are believed
to be the building blocks for certain nanoscale machines and motors
\cite{blanco}, so-called hermaphrodite molecules have been shown to perform
linear relative motion (``contraction and stretching'') 
\cite{jimenez},
and pirouetting molecules have been synthesised \cite{raehm}. (In part,
the inspiration for these new kinds of {\em designer molecules\/}
comes from the biological machinery contained in cellular systems
\cite{alberts}.) Moreover, topological molecules are thought to become
components for molecular electronics switching devices in memory and
computing applications \cite{pease,lehn1}. 
These molecular machines are usually of lower molecular weight, and their
behaviour is essentially energy-dominated in the sense that their
conformations and dynamical properties are governed by external and thermal
activation in an energy landscape. The understanding of the physical
properties and the theoretical modelling of such designer molecules and their
natural biological counterparts has increasingly gained momentum, and the
stage is already set for the next generation of applications
[3-9,\,11-18].
%

In this work, we introduce some basic
concepts for functional molecules whose driving force is entropic rather
than energetic. These molecules will be of higher molecular weight 
(hundred monomers or above)
in order to provide sufficient degrees of freedom such that entropic
effects can determine the behaviour of the molecule.
The potential for such {\em entropy-driven functional molecules\/} 
can be anticipated from the classical Gibbs Free energy
\begin{equation} \label{free}
F = U - T S \, ;
\end{equation}
in functional molecules, $F$ is minimised mainly by variation of the 
internal energy $U$ representing the shape of the energy landscape of 
the functional unit. Here, we propose new types of molecules for which 
$F$ is minimised by variations of the entropy $S$, while the energies 
and chemical bondings are left unchanged. The entropy-functional units 
of such molecules can be specifically controlled by external parameters 
like temperature, light flashes, or other electromagnetic fields
\cite{lehn,dietrich}. We note that DNA is already being studied as
a macromolecular prototype
building block for molecular machines \cite{yurke}.

\begin{figure}
\unitlength=1cm
\begin{picture}(8,4.2)
\put(-1.5,-6.2){\includegraphics{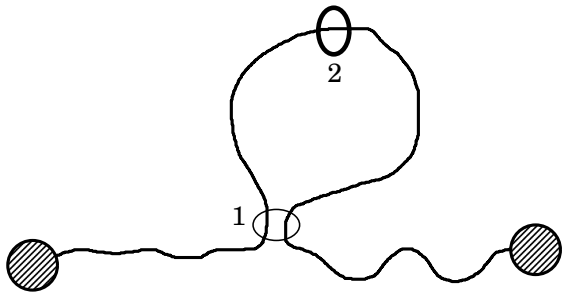}} 
\end{picture}
\caption{Linear polymer chain with stoppers at both ends. A permanent
loop is formed by the slip-link, 1, holding two parts of the chain close 
together. On the loop, a sliding ring, 2, is placed.
\label{fig1}}
\end{figure}

To be more specific, and to gain some intuition about entropic effects
in molecules,
in Fig.\,\ref{fig1} we depict a simple entropy-driven functional 
molecule, which consists of a linear, flexible polymer chain. The chain
is constrained by a cyclic molecule which acts as a {\em slip-link\/},
i.e., it forces two given chain monomers close together such that two 
parts of the chain can freely slide through it
\cite{ball,mehadokaka,mehadokaka1}. Both ends of the chain 
are capped with large groups which act as {\em stoppers\/} to capture 
the slip-link. In addition, the loop threads another cyclic molecule, 
a {\em sliding ring\/} \cite{mehadokaka,mehadokaka1}, which represents
its functional unit. In terms
of the arc length of the loop, the sliding ring is subject to a periodic 
potential, the periodicity and shape of which are determined by the spatial 
extension and the specific chemical composition, respectively, of the 
chain monomers. 

Suppose that, initially, the system is coupled to a heat bath with 
temperature $T_0$ at which the sliding ring is 
trapped in one of the valleys of this periodic potential along the
loop, such that it is immovably placed on it. 
(The sliding ring merely acts to prevent the loop from escaping through 
the slip-link by configurational fluctuations of the polymer chain.)
In this case, the free energy (\ref{free}) of the system is minimised
by maximising its entropy $S$; we have shown in
Ref.~\cite{mehadokaka,mehadokaka1} 
that, accordingly,  the probability density function to find the loop 
with arc length $\ell$ is given by
\begin{equation}
p_0(\ell)={N}\ell^{-3\nu+\sigma_4} \propto \ell^{-2.24} \, \, ,
\end{equation}
where ${N}$ is a normalisation factor, $\nu\approx 0.588$ is the
Flory exponent, and $\sigma_4\approx-0.48$ is a topological exponent
which characterises a vertex at which four segments of a flexible,
self-avoiding chain are linked together \cite{Dup86}. 
The average loop size is given by
\begin{equation}
\langle \ell \rangle_0 = \int_a^{L} d\ell \, \ell \, p_0(\ell) \, \sim a \, ,
\end{equation}
where $L$ is the arc length of the polymer chain and the length 
$a$ represents the smallest possible size of the loop as given by 
the actual design ($a$ is of the order of the monomer size). 
Thus, the loop is entropically favoured to be {\em tight\/}, 
and the whole compound closely behaves like a linear chain 
of length $L$.

A completely different picture results if the sliding ring
is activated, e.g., by increasing the temperature $T$ sufficiently
above $T_0$ or by the influence of external electromagnetic fields, 
such that it can overcome the potential barriers of the periodic 
potential along the loop, and thus can freely slide on it. 
Since for each specific 
configuration of the polymer chain with loop length $\ell$ the sliding 
ring can now occupy $\ell$ different positions along the loop, the 
probability density function is modified to
\begin{equation}
p_1(\ell) = \ell \, p_0(\ell) \propto \ell^{-1.24} \, \, ,
\end{equation}
which implies
\begin{equation}
\langle \ell \rangle_1 = \int_a^{L} d\ell \, \ell \, p_1(\ell) 
\, \sim a^{0.24} \, L^{0.76} \, ,
\end{equation}

\begin{figure}
\unitlength=1cm
\begin{picture}(8,6)
\put(-1.5,-4.6){\includegraphics{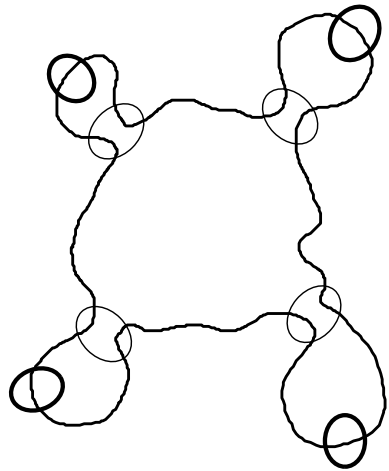}} 
\end{picture}
\caption{Central ring polymer from which a number of slip-links separate
off fringe loops. Within the latter, sliding rings are placed. If the
sliding rings are immobile, the central loop is the overall governing
structure. If the sliding rings are activated and 
become mobile, the fringe loops tend to be large, which leads to a decrease 
of the mean extension of the compound.
\label{fig2}}
\end{figure}
i.e., the additional degrees of freedom of the activated sliding 
ring result in the fact that the loop is not tight but grows with $L$. 
This entropic effect can
be further enhanced by placing not only one, but two or more sliding 
rings on the loop, since $p_n(\ell) = \ell^n \, p_0(\ell)$
and $\langle \ell \rangle_n \sim L$ for $n \ge 2$ activated 
sliding rings. For several sliding rings, the loop will in fact be 
the dominating feature, i.e., the compound behaves more and more like 
a ring polymer of length $L$ as sliding rings are added. 
Thus, for a melt or solution of such molecules, the externally 
controllable activation of the sliding ring(s) leads to a topological, 
entropy-driven transition from linear chain to ring polymer.

The above statements are valid for long chains. We have obtained 
that chains with one loop can be regarded long for about 100 or 
more statistically independent units. If the chain
is shorter, finite size effects come into play and diminish the degrees of
freedom of the sliding rings, and therefore the influence of entropy.
However, this might even be desirable if entropy-functionality is expected
to change the properties of a system only slightly. In the above example,
a shorter chain might exhibit only slightly increased entropy-swelling of
the loop.

Similar entropic effects govern the configuration depicted in 
Fig.\,\ref{fig2}, in which a number of slip-links are placed along a 
ring polymer. Within each of the fringe loops, additional sliding 
rings are placed. If, initially, the sliding 
rings are immobile, the central loop consumes almost the entire length 
of the polymer ring. Conversely,
if the sliding rings are activated, the fringe loops are 
entropically favoured to be large, which should decrease the mean 
extension of the 
entire compound drastically. It would be interesting to create melts 
or gels of such systems; as the single compound (for immobile sliding-rings)
closely behaves like a ring 
polymer, the concatenation of a number of such compounds should 
require no new chemistry.

\begin{figure}
\unitlength=1cm
\begin{picture}(8,5.5)
\put(-1.6,-5.5){\includegraphics{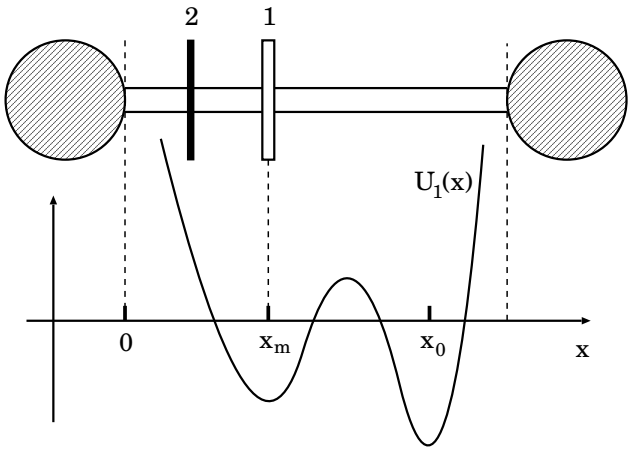}} 
\end{picture}
\caption{Modified rotaxane molecule. The usual rotaxane ring, 1,
and the sliding ring, 2, are placed along the molecule strand,
between the stopper balls (e.g., C$_{60}$ \protect\cite{dietrich}).
Without activated sliding ring, the rotaxane ring is subject to
a potential landscape $U_1(x)$ (here sketched schematically with a stable
position $x_0$ and a metastable position $x_m$). If the sliding
ring is activated, it exerts an additional entropic force on the
rotaxane ring, which tends to shift it to the right.
\label{fig3}}
\end{figure}

To see how energetic and entropic contributions compete with each other,
consider the idealised rotaxane depicted in Fig.\,\ref{fig3}. Typically,
a rotaxane consists of a ring, 1,  on a backbone strand and is subject to a 
potential landscape
$U_1(x)$ with a stable minimum at a position $x_0$ and a metastable 
minimum at another position $x_m$.
A laser pulse can be used to move the rotaxane ring from $x_0$ to $x_m$ 
before it relaxes back to $x_0$. Since the potential barrier(s) between
$x_m$ and $x_0$ can be high, the refractory times can be fairly long
(e.g., of the order of days) \cite{lehn,dietrich,blanco,metz}.
In a modified rotaxane molecule, one could put an additional
sliding ring which, if activated, can freely slide along the backbone 
of the rotaxane between the stopper and the rotaxane ring,
like a particle of a one dimensional ideal gas.
The free energy (\ref{free}) of the system
for a fixed position $x$ of the rotaxane ring becomes
\begin{equation} \label{rot}
F(x) = U_1(x) - k_B T \log(x/a) \, ,
\end{equation}
where $k_B$ is Boltzmann's constant and $x/a$ is the number of
sites the sliding ring, 2, can take on. On the rotaxane 
ring therefore acts the effective force
\begin{equation}
K(x) = - \, F'(x) = - \, U_1'(x) + k_B T /x \, \, ,
\end{equation}
where the entropic contribution $k_BT/x$ tends to shift the rotaxane ring
to the right. Thus, entropy-functional units could be used, e.g., 
to accelerate the often extremely long refractory times in rotaxanes.
Conversely, if placed on the opposite side of the metastable rotaxane
ring position and activated, the sliding ring could increase the 
refractory time and thus lead to a stabilisation of 
the metastable position of the rotaxane ring for the 
purpose of information storage.
We note that it should be possible to synthesise long rod-like axles 
for entropy-functional rotaxane molecules by using translationally 
invariant homopolymer helices \cite{poland}.

\begin{figure}
\unitlength=1cm
\begin{picture}(8,3)
\put(-0.3,0.0){\includegraphics{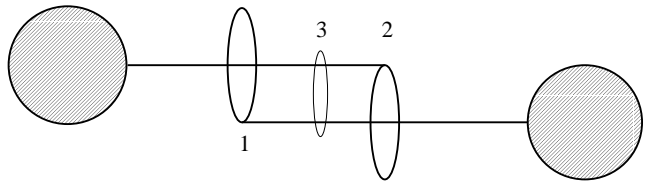}} 
\end{picture}
\caption{``Molecular muscle'' consisting of two interlocked rings 
1 and 2 with attached rod-like molecules. Within this structure, sliding
rings, 3, can be placed, which, if activated, tend to contract the muscle
by entropic forces.}
\label{fig4}
\end{figure}

Finally, consider the nanomechanical device depicted in Fig.\,\ref{fig4}.
According to the arrangement of the sliding rings 1 and 2, 
this compound exhibits the so far unique

\newpage

\noindent
feature of a molecule that it can slide {\em laterally\/}.
Suggested as precursors of molecular muscles \cite{jimenez}, this compound
could be propelled with internal entropy-motors, which entropically
adjust the elongation of the muscle. In the configuration shown in 
Fig.\,\ref{fig4}, the sliding ring 3 creates, if activated, an 
entropic force which tends to contract the ``muscle''; 
at $T = 300 \, {\mbox K}$ and on a typical scale $x = 10 \, \mbox{nm}$, 
the entropic force $k_B T / x$ 
is of the order of pN, and thus comparable to the force created
in biological muscle cells \cite{gittes}. Molecular muscles of such a make
can be viewed as the nano-counterpart of macroscopic muscle models proposed
by de Gennes \cite{degennesm}, in which the contraction is based on 
the entropy difference between the isotropic and nematic phases in liquid 
crystalline elastomer films \cite{thomsen}.

Entropy has so far been disregarded in the design of functional molecules,
although the basis for this, i.e., the mechanically interlocked state of
topological molecules, has been achieved for some time, and prototype
molecules with sliding rings have been synthesised \cite{dietrich}.
Entropy-functional units can be specifically controlled by external
parameters, e.g., temperature and electromagnetic fields.
Functional behaviour such as controlled transition from linear chain
to ring polymer, swelling/de-swelling, switching in rotaxane-like
compounds, and molecular motion (muscle contraction) could thus be
achieved without changing the chemical structure of the involved
compounds (the latter is done in designed biomolecular motors
\cite{montemagno,soong,yurke}). 
As another possible application, one might speculate 
whether the DNA helix-coil transition \cite{poland}
in multiplication setups could be
facilitated in the presence of pre-ring molecules which in vitro attach
to an opened loop of the double strand and close, creating an entropy 
pressure which tends to open up the vicinal parts of the DNA which are 
still in the helix state. Finally, considering molecular motors,
it would be interesting to design an externally controllable, 
purely entropy-driven rotating nanomotor. 

We thank M. Schick for helpful discussions.
We acknowledge financial support from the Deutsche Forschungsgemeinschaft
(DFG). A.H. also acknowledges support from the National Science Foundation
under grant No. 6892372 and from the Engineering and 
Physical Sciences Research Council through grant GR/J78327.

\end{document}